\documentclass[prd,twocolumn]{revtex4} % article
\usepackage{amsmath}
\usepackage{amssymb}

\begin{document}

\title{Extended axion electrodynamics: Anomalous dynamo-optical \\
response induced by  gravitational pp-waves}

\author{Alexander B. Balakin} 
\email{Alexander.Balakin@kpfu.ru}  
\affiliation{Kazan Federal University, Institute of Physics, Kremlevskaya
str., 18, Kazan, 420008, Russia}

\author{Timur Yu. Alpin}
\email{Timur.Alpin@kpfu.ru}
\affiliation{Kazan Federal University, Institute of Physics, Kremlevskaya
str., 18, Kazan, 420008, Russia}

\begin{abstract}
We extend the Einstein-Maxwell-axion theory including
into the Lagrangian cross-terms of the dynamo-optical type, which
are quadratic in the Maxwell tensor, linear in the covariant
derivative of the macroscopic velocity four-vector, and linear in
the pseudoscalar (axion) field or its gradient four-vector. We
classify the new terms with respect to irreducible elements of the
covariant derivative of the macroscopic velocity four-vector of
the electromagnetically active medium: the expansion scalar,
acceleration four-vector, shear and vorticity tensors. Master
equations of the extended axion electrodynamics are used for the
description of the response of an axionically active
electrodynamic system, induced by  a pp-wave gravitational
background. We show that this response has a critical character,
i.e., the electric and magnetic fields, dynamo-optically coupled
to the axions, grow anomalously under the influence of the
external pp-wave gravitational field. 
\end{abstract}

\maketitle

\section{Introduction}

Axion electrodynamics as an extension of the Faraday - Maxwell
electromagnetic theory is based on the prediction of axions,
massive pseudo-Goldstone bosons \cite{PQ,Weinberg,Wilczek}. The
first discussion concerning the pseudoscalar-photon interaction
appeared in \cite{WTN77}; however, the interest to axion
electrodynamics has grown significantly later, after publication
of the paper \cite{Sikivie}. Important aspects of the axion theory
and of its astrophysical and cosmological applications can be
found, e.g., in reviews and book \cite{ADM1,ADM2,ADM3,ADM4,ADM5}.
Results of experimental investigations of the axion-photon
interactions are published, e.g., in
\cite{PVLAS2,GammeV,CAST2,OSQAR,QA,BMV2}).

The axion electrodynamics can be indicated as the standard one,
when the Lagrangian contains only one cross-term $\frac14 \phi
F^{*}_{mn}F^{mn}$, in which the product of the pseudoscalar
(axion) field $\phi$ and of dual Maxwell (pseudo)tensor behaves as
a true tensor. In that theory axion-induced phenomena can be
visualized, when the axion field $\phi$ has non-vanishing gradient
four-vector $\nabla_k \phi \neq 0$; respectively, the invariant $I
{=} g^{ik} \nabla_i \phi \nabla_k \phi$ can be positive, negative
or equal to zero. The last case relates to models with pp-wave
symmetry, for which the axion field depends on the retarded time
only. Here we focus just on this model and consider the pp-wave
gravitational background as a scene for evolution of the
electromagnetic field in a non-uniformly moving axionically active
medium.

In the papers \cite{B1,B2,B3,B4,B5,B6,B7} we considered extensions
of the axion electrodynamics, keeping in mind that even if the
axion field is initially constant, external gravitational and
electromagnetic fields are able to activate frozen axion-photon
couplings in course of evolution of the corresponding physical
system. In particular, we considered a non-minimal axion-photon
coupling activated by gravitational waves \cite{B1};
non-stationary optical activity and gradient-type models of the
axion-photon coupling in a cosmological context \cite{B2,B3};
fingerprints of dark matter axions in the terrestrial electric and
magnetic field variations \cite{B4}; electromagnetic waves in an
axion-active plasma \cite{B5,B6}; axionically induced anomalous
behavior of the electromagnetic response to the gravitational wave
action \cite{B7}.

In this work we consider the so-called {\it dynamo-optical
extension} of the axion electrodynamics. This term was introduced
in \cite{LLP} to describe the influence of a non-uniform
(irregular) motion of a physical system on its electromagnetic
response. The generalization of dynamo-optical type of the
Einstein-Maxwell theory was carried out in \cite{AB2006}. In this
paper we add a new element into the theory, namely, the
pseudoscalar (axion) field, thus providing the dynamo-optical
extension of the axion electrodynamics, which is a distinctive
part of the Einstein-Maxwell-axion theory.

\section{The model}

\subsection{Action functional and standard definitions}

Master equations of the extended axion electrodynamics form the
sub-set of the total system of equations of the Einstein - Maxwell
- axion model (see, e.g., \cite{B1,B7}); they can be obtained by
variation of the action functional
\begin{equation}
S{=}\int d^4x \sqrt{{-}g} \left[ \frac{R}{2\kappa} {+}L_{({\mathrm
m})} {+}L_{({\rm EM})} {+}L_{({\mathrm A})} \right] \label{act}
\end{equation}
with respect to the electromagnetic potential four-vector $A_i$,
and dimensionless pseudoscalar (axion) field $\phi$, respectively.
Equations of the gravity field, as a result of variation of this
action functional with respect to space-time metric $g_{ik}$, are
presented in \cite{B6} and we do not consider them in this paper.
As usual, $g$ is the determinant of the metric; $R$ is the Ricci
scalar; $\kappa {=} \frac{8 \pi G}{c^4}$ is the Einstein constant.
The Lagrangian of the axion field is of the form
\begin{equation}
L_{({\rm A})} {=} \frac12 \Psi_{0}^2 \left[m^2_{({\rm A})}\phi
^{2} {+}V(\phi^2) {-} g^{mn}\nabla_{m}\phi \nabla_{n}\phi \right]
, \label{LA}
\end{equation}
where $\nabla_{m}$ is a covariant derivative; the parameter
$\frac{1}{\Psi_{0}}$ is a coupling constant of the axion-photon
interaction; the term $m_{({\rm A})} {=} \frac{c}{\hbar} m_{({\rm
axion})}$ is a re-scaled axion mass $m_{({\rm axion})}$; $\hbar$
is the Planck constant. The total Lagrangian of the
electromagnetic field in the axionically active medium, $L_{({\rm
EM})}$, is considered to be quadratic with respect to the Maxwell
tensor $F_{ik} \equiv \nabla_i A_{k} {-} \nabla_k A_{i}$. This
Lagrangian includes the dual tensor $F^{*mn} = \frac{1}{2}
\epsilon^{mnpq}F_{pq}$, where, as usual, $\epsilon^{mnpq} \equiv
\frac{1}{\sqrt{{-}g}} E^{mnpq}$ is the Levi-Civita tensor,
$E^{mnpq}$ is the skew-symmetric Levi-Civita symbol with
$E^{0123}{=}1$. The dual Maxwell tensor satisfies the condition
$\nabla_{k} F^{*ik} {=}0$, which is treated as a sub-set of
electrodynamic equations.

Keeping in mind the necessity of a phenomenological decomposition
of the extended Lagrangian $L_{({\rm EM})}$ into irreducible
parts, we use the following standard representation of the tensor
$F^{ik}$:
\begin{equation}
F^{ik} = E^i U^k -  E^k U^i - \epsilon^{ikmn}B_m U_n  \,.
\label{decompF}
\end{equation}
Here $U^i$ is the macroscopic velocity four-vector; we use the
Landau-Lifshitz definition of $U^i$ considering it as a time-like
eigen-vector of the stress-energy tensor of the matter (see, e.g.,
\cite{B7} for details). The electric field four-vector, $E^i
\equiv F^{ik} U_k$, and the magnetic induction four-vector, $B_i
\equiv F^{*}_{ik} U^k$, are, clearly, orthogonal to the velocity
four-vector $U^i$.

In order to classify irreducible terms describing interactions of
the dynamo-optical type, we use in the decomposition of the
Lagrangian $L_{({\rm EM})}$ the following standard representation
of the covariant derivative of the velocity four-vector:
\begin{equation}
\nabla_i U_k = U_i D U_k + \sigma_{ik} + \omega_{ik} + \frac{1}{3}
\Delta_{ik} \Theta \,. \label{decompDU}
\end{equation}
Here $DU_k \equiv U^i \nabla_i U_k$ is the medium acceleration
four-vector; $ \Theta \equiv \nabla_m U^m$ is the expansion
scalar; the traceless symmetric shear tensor $\sigma_{ik}$, the
skew-symmetric vorticity tensor $\omega_{ik}$, and the projector
$\Delta_{ik}$ are given by the formulas
$$
\sigma_{ik} \equiv \Delta_{(i}^{p} \Delta_{k)}^{q}\nabla_p U_q {-}
\frac{1}{3} \Delta_{ik} \Theta \,, \quad
$$
\begin{equation}
\omega_{ik} \equiv \Delta_{[i}^{p} \Delta_{k]}^{q} \nabla_p U_q
\,, \quad \Delta_{ik} = g_{ik} {-} U_iU_k \,. \label{decompSO}
\end{equation}
The symbols $(ik)$ and $[ik]$ define the operations of
symmetrization and skew-symmetrization, respectively. Also, we use
two auxiliary tensors: the angular velocity (pseudo) four-vector
$\omega_i$, and the skew-symmetric tensor $\Omega_{pq}$ given by
\begin{equation}
\omega_i \equiv - \omega^{*}_{ik}U^k \,, \quad \Omega_{pq} \equiv
U_{[p} DU_{q]} \,. \label{decompSO2}
\end{equation}
We consider the quantities
$\phi$,$\nabla_i\phi$,$E^i$,$B^k$,$U^i$,$DU_i$, $\Theta$,
$\sigma_{ik}$, $\omega_{ik}$ to be basic elements of the
decomposition of the electromagnetic field Lagrangian.

\subsection{Decomposition of the term $L_{({\rm EM})}$}

We divide the electromagnetic field Lagrangian $L_{({\rm EM})}$
into seven parts
\begin{equation}
L_{({\rm EM})}= L_{0}+L_{1}+L_{2}+L_{3}+L_{4}+L_{5}+L_{6} \,.
\label{decompSO3}
\end{equation}
For our purposes it is convenient to write the term $L_{0}$ in the
following  form
\begin{equation}
L_{0} = \frac{1}{2\mu} \left(n^2 E^iE_i - B^iB_i \right)+\phi B^i
E_i  \,, \label{decompSO4}
\end{equation}
using the $(E,B)$-representation. Here $n^2 \equiv \varepsilon
\mu$, i.e., $n$ is the refraction index of the medium with
dielectric and magnetic permittivities $\varepsilon$ and $\mu$,
respectively. When $\varepsilon{=}1$, $\mu{=}1$, the term
(\ref{decompSO4}) is of the form $\frac14 \left(F^{mn}F_{mn} {+}
\phi F^{mn}F^{*}_{mn}\right)$, thus recovering the Lagrangian of
the standard axion electrodynamics.

The term $L_{1}$ does not contain the axion field $\phi$, however,
in contrast to (\ref{decompSO4}), it is linear in the covariant
derivative of the velocity four-vector and contains six
irreducible scalar terms equipped by six phenomenological
constants $\lambda_{11}$, ..., $\lambda_{16}$ \cite{AB2006}:
$$
L_{1} = \frac{1}{4}  \Theta  \left( \lambda_{11} E_{k}E^{k} {+}
\lambda_{12}B_k B^{k}\right)+
$$
$$
 \ \ \ \ \ \ \ \ \ {+}\frac{1}{4}  \sigma^{km}
\left(\lambda_{13}E_{k}E_{m}{+} \lambda_{14}B_k B_{m}\right) +
$$
\begin{equation}
\ \ \ \ \ \ \ \ \ \ \ \ \ \ \ + \frac{1}{4} \epsilon^{ikmn} E_i
B_k \left(\lambda_{15}\Omega_{mn} {+}\lambda_{16}  \omega_{mn}
\right)\,. \label{decompSO5}
\end{equation}
Terms linear in the pseudoscalar field $\phi$ and in the
convective derivative $D\phi$ contain pseudovector $B^i$
providing $L_{2}$ and $L_{3}$ to be pure scalars:
\begin{equation}
L_{2} {=} \frac{\phi}{4}E_m B_n \left(\lambda_{21} \Theta g^{mn}
{+} \lambda_{22}\sigma^{mn} {+} \lambda_{23}\omega^{mn} \right),
\label{decompSO61}
\end{equation}
\begin{equation}
L_{3} {=} \frac{D\phi}{4}E_m B_n \left(\lambda_{31}\Theta g^{mn}
{+} \lambda_{32} \sigma^{mn} {+} \lambda_{33} \omega^{mn} \right)
. \label{decompSO6}
\end{equation}
All terms linear in the spatial gradient of the pseudoscalar
field, $\Delta^k_i \nabla_k \phi$, can be reduced to the set of
three scalars $L_{4}$, $L_{5}$, $L_{6}$. First, we construct
scalars linear in the vorticity tensor; it is convenient to use in
this case the pseudovector $\omega_i$ guaranteeing the product
$\omega_{m} \nabla_{n} \phi $ to be a pure tensor:
$$
L_{4} = \frac{1}{4} \omega_{(m} \nabla_{n)} \phi \left[
\Delta^{mn}\left( \lambda_{41} E_k E^k + \lambda_{42}B_k B^k
\right) + \right.
$$
\begin{equation}
\left. \ \ \ \ \ \ + \left(\lambda_{43} E^m E^n + \lambda_{44} B^m
B^n\right) \right]\,. \label{decompSO7}
\end{equation}
Second, we list the terms linear in the acceleration four-vector
$DU_k$ (terms with $\phi$ can not appear):
$$
L_{5} = \frac{1}{4} \nabla_{n}\phi \ DU_{m}  \left( \lambda_{51}
\Delta^{mn} E^k B_k + \right.
$$
\begin{equation}
\left. \ \ \ \ \ \ \ \ \ \ \ \ \  \ \ \ + \lambda_{52} E^n B^m +
\lambda_{53} E^m B^n \right)\,. \label{decompSO8}
\end{equation}
Third, there are two terms linear in the shear tensor
\begin{equation}
L_{6} {=} \frac{1}{4}   \eta^{nmp} \sigma_{mk}\nabla_{n}\phi
\left(\lambda_{61}
 E^k E_p {+} \lambda_{62}B^k B_p
\right). \label{decompSO9}
\end{equation}
Our statement is that the decomposition presented above is
irreducible, and the number of independent coupling constants
$\lambda_{11}$, ... ,$\lambda_{62}$, appeared in front of listed
terms, is twenty one. Keeping in mind symmetry motives, further we
reduce the number of these parameters using the following
relationships:
$$
\lambda_{13}{=}{-}\lambda_{12}\,, \quad
\lambda_{13}{=}\lambda_{14}\,, \quad
\lambda_{21}{=}\frac{1}{3}\lambda_{22}\,, \quad
\lambda_{31}{=}\frac{1}{3}\lambda_{32}\,,
$$
\begin{equation}
\lambda_{42}{=}{-}\lambda_{41}\,, \quad
\lambda_{44}{=}\lambda_{33}\,, \quad \lambda_{62}{=}\lambda_{61}
\,. \label{decompSO91}
\end{equation}
 Also, we put $\lambda_{16}{=}0$,
since the corresponding term does not enter the electrodynamic
equations due to their specific structure.

\subsection{Master equations}

\subsubsection{A pp-wave gravitational background}

We consider test electromagnetic and pseudoscalar fields in the
pp-wave gravitational background with the line element
\begin{equation}
\mbox{d}s^{2} = 2\mbox{d}u\mbox{d}v {-} L^{2} \left[e^{2
\beta}(\mbox{d}x^2)^2 {+} e^{{-}2\beta} (\mbox{d}x^3)^2 \right]
\,, \label{ppmetric}
\end{equation}
where $u {=} \frac{ct {-} x^1}{\sqrt{2}}$ is the retarded time and
$v {=} \frac{ct {+} x^1}{\sqrt{2}}$ is the advanced time. For such
model the background factor $L(u)$ satisfies the requirements
\begin{equation}
L^{\prime \prime} {+} \left(\beta^{\prime}\right)^2 L = 0 \,,
\quad L(0){=}1 \,, \quad L^{\prime}(0) {=} 0 \,, \label{Lpp}
\end{equation}
where the prime denotes the derivative with respect to the
retarded time $u$, and $\beta(u)$ is arbitrary function of the
retarded time with initial value $\beta(0){=}0$. We consider the
medium to be at rest, and the velocity four-vector to be of the
form $U_u{=} U_v{=}\frac{1}{\sqrt{2}}$, $U_2{=}U_3{=}0$. For this
case $DU_i{=}0$, $\omega_{ik}{=}0$, $\Theta {=}\frac{\sqrt2
L^{\prime}}{L}$,
\begin{equation}
\sigma^k_i  = \frac{\Theta}{2}  \left(\frac13 \Delta_i^k {-}
\delta_i^1 \delta^k_1 \right) {+} \frac{\beta^{\prime}}{\sqrt2}
\left(\delta_i^2 \delta^k_2 {-} \delta_i^3 \delta^k_3 \right)\,.
\label{GW002}
\end{equation}
We assume that the electromagnetic and axion fields inherit the
pp-wave symmetry of the gravitational background, thus, the
unknown functions depend on $u$ only: $E^i(u)$, $B^k(u)$,
$\phi(u)$ (see \cite{B7}).

\subsubsection{Electrodynamic equations}

Electrodynamic equations have the standard form
\begin{equation}
\nabla_k H^{ik} = 0 \,, \quad \nabla_k F^{*ik} = 0 \,, \label{ED1}
\end{equation}
where the excitation tensor can be now written as
\begin{equation}
H^{ik} = U^n \left[\delta^{ik}_{mn} \frac{\partial L_{({\rm
EM})}}{\partial E_m} + \epsilon^{ik}_{\ \ mn} \frac{\partial
L_{({\rm EM})}}{\partial B_m} \right] \,, \label{ED2}
\end{equation}
using the explicit $(E,B)$ - representation of the scalar
$L_{({\rm EM})}$ (\ref{decompSO3})-(\ref{decompSO9}). This
procedure is routine, and we omit details of the $H^{ik}$
decomposition. When unknown functions depend on retarded time
only, integration of (\ref{ED1}) gives six integrals
\begin{equation}
L^2 H^{iu}(u) {=} H^{iu}(0) \,, \quad L^2 F^{*iu}(u) {=}
F^{*iu}(0) \,, \label{ED3}
\end{equation}
($i{=}v,x^2,x^3$). Three of them do not contain $\phi(u)$:
$$
B_{v}(u) {=} \frac{B_{v}(0)}{L^2(u)}, \ B_{2}(u) {=} e^{2\beta}
\left[B_{2}(0) {+} E_{3}(u) {-} E_{3}(0)\right],
$$
\begin{equation}
 B_{3} = e^{-2\beta} \left[B_{3}(0) +E_{2}(0)-E_{2}(u)\right]\,.
\label{BB1}
\end{equation}
The longitudinal integral (for $i{=}v$)yields
\begin{equation}
E_{v}(u)  {=} \frac{\varepsilon E_{v}(0)-
B_{v}(0)[\phi(u)-\phi(0)]}{L^2\left[\varepsilon + \frac16
\Theta(u)(3\lambda_{11}{-}\lambda_{13})\right]}\,. \label{BB19}
\end{equation}
Two remaining integrals contain both $\phi(u)$ and
$\phi^{\prime}(u)$; using (\ref{BB1}) they can be written in the
form:
$$
2{\cal A}(u) e^{- 2\beta}E_{2}(u) + 3 \beta^{\prime}{\cal B}(u)
E_{3}(u) = {\cal J}_{2}(u)\,,
$$
\begin{equation}
\ \ \ 3 \beta^{\prime}{\cal B}(u) E_{2}(u) + 2{\cal A}(u)
e^{2\beta} E_{3}(u) = {\cal J}_{3}(u)\,,
 \label{BB2}
\end{equation}
where the auxiliary functions are defined as follows:
$$
{\cal A}(u) = \lambda_{13} \Theta + 6\left(\varepsilon -
\frac{1}{\mu} \right) \,,
$$
\begin{equation}
\ \ \ \ \ \ \ \ \ \ \ {\cal B}(u) = \lambda_{32} \phi^{\prime}+
\sqrt2\lambda_{22} \phi \,.
 \label{BB3}
\end{equation}
The source-like term ${\cal J}_{2}(u)$ in (\ref{BB2}) is of the
form
$$
{\cal J}_{2}(u) =
\frac{3}{2\sqrt2}\left[E_3(0){-}B_2(0)\right]\left\{6
\lambda_{22}\beta^{\prime}\phi {+} \Theta {\cal B} {+}\right.
$$
$$
\left. {+} \sqrt2 \beta^{\prime}
\phi^{\prime}(\lambda_{32}{-}2\lambda_{61}) {+} 48\sqrt2
[\phi-\phi(0)] \right\} {+}
$$
$$
{+}\left[B_3(0){+}E_2(0) \right]\left\{e^{{-}2\beta}\left[\Theta
(\lambda_{13}{-}6 \lambda_{11}){-}3\sqrt2 \lambda_{13}
\beta^{\prime} \right] \right.
$$
\begin{equation}
\left. {+}\frac{12}{\mu}\left(1{-}e^{{-}2\beta} \right) \right\}
{+} 12 E_2(0)\left(\varepsilon - \frac{1}{\mu} \right)
\,.\label{BB4}
\end{equation}
The term ${\cal J}_{3}$ can be obtained from (\ref{BB4}) by
replacements $\beta \to {-}\beta$, $E_2(0)\to E_3(0)$, $B_2(0)\to
{-}B_3(0)$.

\subsubsection{Equation of the axion field evolution}

Variation of the action functional (\ref{act}) with respect to the
axion field $\phi$ gives the equation
\begin{equation}
\left[ \Box {+} m_{({\rm A})}^{2} {+} V'(\phi^2) \right] \phi =
{-} \frac{1}{\Psi^2_{0}}\left(B_k E^k {-} {\cal J} \right) \,.
\label{AX1}
\end{equation}
When ${\cal J}{=}0$, we deal with standard equation for the
pseudoscalar field; using explicit representation of $L_{2},...,
L_{6}$ (see (\ref{decompSO61})-(\ref{decompSO9})) we can write the
dynamo-optical source ${\cal J}$ in the following compact form:
$$
{\cal J}= 2 \left\{ {-} \left(\frac{L_2}{\phi}\right) + (\Theta
{+} D)\left( \frac{L_3}{D\phi}\right){+} \right.
$$
\begin{equation}
\left. \ \ \ \ \ \ \ \ \ \  \ \ +\nabla_n \left[
\frac{\partial}{\partial (\nabla_n \phi)}\left(L_4{+}L_5{+}L_6
\right)\right] \right\} \,. \label{AX2}
\end{equation}
Clearly, this term is quadratic in the components of the
electromagnetic field, $E^i$ and $B^k$; it is up to second order
with respect to irreducible elements of the decomposition of the
covariant derivative of the velocity four vector, $\Theta$,
$DU_k$, $\sigma_{mn}$, $\omega_{mn}$, and contains second
covariant derivatives $\nabla_s \nabla_m U_n$.

\subsubsection{Initial state}

When the gravitational wave is absent ($u<0$), the pseudoscalar,
electric and magnetic fields are considered to be constant
($\phi(0)$, $E_i(0)$, $B_i(0)$, respectively), and the system as a
whole to be at rest. Clearly, at $u{=}0$ the equations (\ref{BB1})
- (\ref{BB2}) with (\ref{BB3}) and (\ref{BB4}) convert into
identities for arbitrary $\phi(0)$, $E_i(0)$, $B_i(0)$. The
initial value $\phi(0)$ can be found from the reduced equation
(\ref{AX1}) with ${\cal J}{=}0$:
\begin{equation}
\left[m_{({\rm A})}^{2} {+} V'(\phi^2(0)) \right] \phi(0) {=} {-}
\frac{1}{\Psi^2_{0}}B_k(0) E^k(0), \label{AX11}
\end{equation}
the values $E_i(0)$, $B_i(0)$ being arbitrary constants.

\section{Anomalous behavior of the electromagnetic response}

The presented above set of master equations of the axion
electrodynamics in the pp-wave gravitational background (see
(\ref{BB1})-(\ref{AX2})) is a coupled system of nonlinear
equations; search for exact solution of this system is a
sophisticated but very interesting problem. In this sense the
exact solutions discussed in \cite{B7} relate to the case of
absence of the dynamo-optic phenomena, nevertheless, that results
let us expect the formulated dynamo-optical problem to be solved
in the nearest future with some special conditions for the
coupling constants. Here we restrict our-selves by analysis of a
truncated model, which, nevertheless, displays the main new
feature: the anomalous character of the axion-photon coupling in
the medium moving non-uniformly.

\subsection{Longitudinal electromagnetic fields}

We use the term {\it longitudinal}, when the initial electric
and/or magnetic fields are directed along the axis of the
gravitational wave propagation. In this case transversal
components can not be produced, thus, we deal with magnetic and
electric fields given by (\ref{BB1}), (\ref{BB19}), and with the
axion field of the form:
$$
\phi(u){=} \frac{\phi(0)\left[\varepsilon m^2_{({\rm A})}
\Psi^2_0{+}2B^2_{v}(0)\right]}{L^4 m^2_{({\rm A})} \Psi^2_0
\left[\varepsilon {+}\frac16 \Theta
(3\lambda_{11}{-}\lambda_{13})\right]+ 2B^2_{v}(0)} ,
$$
\begin{equation}
\ \phi(0){=} \frac{2 E_v(0)B_{v}(0)}{m^2_{({\rm A})} \Psi^2_0} \,,
\quad (V(\phi^2) \equiv 0) \,. \label{AX24}
\end{equation}
The denominator of $E_{v}(u)$ in (\ref{BB19}) can tend to zero
value at $u\to u^*$, when $\Theta(u^{*}){=}6 \varepsilon
(\lambda_{13} {-}3\lambda_{11})^{{-}1}$, thus providing an
anomalous growth of the longitudinal electric response. The axion
field at this moment, $\phi(u^*)$, remains finite, when
$B_{v}(0)\neq 0$.

\subsection{Transversal electromagnetic fields}

When $E_v(0){=}B_v(0){=}0$, but $E_2(0),...B_3(0)\neq 0$, we deal
with three-dimensional nonlinear system of coupled evolutionary
equations. Let us illustrate the appearance of the response
anomaly for the case of relic dark matter axion domination (see,
e.g., \cite{B3,B4}). This term means that the density of axions,
produced by electromagnetic field, is much smaller that the
density of relic cosmological axions. In such case we consider the
function $\phi(u)$ to be fixed, and face with linear algebraic
system (\ref{BB2}). The corresponding Cramer's determinant
$\Delta(u)$
$$
\Delta(u) = 4 {\cal A}^2(u)- 9 {\beta^{\prime}}^2 {\cal B}^2(u)
$$
takes zero value at the moment $u_{*}$, for which $ {\cal
A}(u_{*}){=} \pm \frac32 \beta^{\prime} {\cal B}(u_{*})$, or in
more details
$$
\Theta(u_{*}) {=}
\frac{6}{\lambda_{13}}\left[\left(\frac{1}{\mu}{-}\varepsilon
\right) {\pm}
\frac{\beta^{\prime}}{4}\left(\lambda_{32}\phi^{\prime}{+} \sqrt2
\lambda_{22}\phi \right) \right]_{|u_{*}}.
$$
This anomaly, appeared in the response of transversal electric and
magnetic fields, is mixed and can be indicated as
axionic-dynamo-optical, in contrast to pure dynamo-optical
longitudinal anomaly in (\ref{BB19}).

\section{Conclusions}

\noindent 1. Dynamo - optical extension of the Einstein - Maxwell
- axion theory is shown to have twenty one coupling constants as a
maximum; the number of essential parameters can be reduced to
thirteen by physical assumptions.

\noindent 2. Longitudinal electric and magnetic fields in an
axionically active medium, evolving in the pp-wave gravitational
background, can possess dynamo - optical anomaly, which describes
an amplification of the electromagnetic response, when
$3\lambda_{11}{-}\lambda_{13} \neq 0$.

\noindent 3. Transversal electric and magnetic fields can display
an anomaly of a new type, which differs from the dynamo-optical
one by the dependence on the axion field.

\section{Acknowledgments}
This work is supported by the Russian Foundation for Basic
Research (Grant No. 14-02-00598).

\end{document}